\documentclass[12pt]{article}
\usepackage{graphicx}
\usepackage{amsmath}
\usepackage{amssymb}
\usepackage{hyperref}
\usepackage[section]{placeins}
\usepackage{authblk}

\begin{document}

\title{Probabilistic Analysis of Event-Mode Experimental Data \\
 
 \small{a.k.a.\\ A Total Noobs Guide to Computational Bayesian Analysis \\ a.k.a. \\ Why We Should Probably Not Use Least Squares Fitting \\ for a Lot of Neutron Experiments}
 }
 
\author[1]{Phillip M. Bentley}
\author[2]{Thomas H.\ Rod}
\affil[1]{European Spallation Source ERIC, P.O. Box 176, 221 00 Lund, Sweden; phil.m.bentley@gmail.com}
\affil[2]{European Spallation Source ERIC, Data Management and Scientific Computing Centre, Asmussens Allé 305, 2800 Kgs. Lyngby, Denmark; thomas.holmrod@ess.eu}


\date{\today}

\maketitle

\begin{abstract}
Neutron and x-ray scattering experiments traditionally rely upon histogrammed data sets, which are analysed using least-squares curve fitting of multiple probability distribution components to quantify separately the various scientific contributions of interest.  The main advantage to these methods is the relative ease of deployment due to their intuitive nature.  Despite great popularity, these methods have known drawbacks, which can cause systematic errors and biases in some common scenarios in this field.  Improvements over the base methods include dynamic optimisation of histogram bin width and the application of modern numerical optimisation methods that have greater stability, but, whilst reduced, the systematic effects carried by this stack nonetheless remain.  In this study, we demonstrate analysis of neutron scattering event data using neither any numerical integration or histogramming steps, nor least squares fitting.  Instead, a numerical Bayesian workflow is applied to each neutron event as it arrives in a data stream.  The main benefits of the new methodology are greater efficiency (i.e. fewer data points required for the same parameter accuracy) and a reduced impact of inherent systematic error in some cases, such as long-tailed distributions.  The main drawbacks are a less intuitive analysis method and an increase in computation time.
\end{abstract}

\section{Introduction}
Historically, in particle-based experiments the accumulation of data events into histograms was performed by hardware.  Each detector event was added to a physical hardware counter, which in turn was read out to an array in computer software.  The readout of the array then provides the dependent variables $y_i$, which are distributed as a probability function $f$ of some independent variable, $x$.  This could be a continuous parameter, e.g. time, distance, but in practice it is digitised onto some grid to create $(x_i,y_i)$ pairs.  Part of the commissioning of the experiment involves calibration of the $x_i$ values for each pixel or histogram ``bin'', $i$.

Notice that some information is lost in the histogramming process, because we are integrating over a range of $x$ values for each bin.  Whilst there is a resolution of the instrument that would be manifest as some variance in the $x$ axis, $\delta x$, there is also a resolution in the histogram itself, $\Delta x = x_{i+1} - x_i$.  For the purposes of this report, we do not complicate matters by concerning ourselves with the optimisation of $\delta x$.  The reason for this is partly because some of the components of $\delta x$ are instrument-specific, but also because, with modern hardware, the computational technology available means that some of the driving practical limitations of some components have become negligible.  Instead, we simply note that it is now possible to record the events as an almost continuous parameter of $x$ down to double-precision floating point resolution and store, for example, the position of the detection event along a wire, or the detection time, since those resolution components can be negligibly small relative to the instrument resolution effects.  Furthermore, the cost of computer storage has reduced considerably over the last few decades.  In neutron scattering instrumentation, this kind of data storage is known as ``event mode'' to distinguish it from histograms.  Typically, a scientist can reconstruct histograms of the data after the experiment is complete depending on the sparsity of the data or the underlying probability distributions $y \sim f(x)$.

As a simple example, in small angle neutron scattering (SANS) one measures intensity as a function of neutron momentum transfer $Q$, defined by:
\begin{equation}
\mathbf{Q} = \mathbf{k}_i - \mathbf{k}_f
\end{equation}
If we assume elastic scattering, then $|\mathbf{k}_i| = |\mathbf{k}_f| = k$, and if we radially average then the direction of $\mathbf{k}$ is integrated out, so only the length of the $\mathbf{k}$ vector is important, so:
\begin{equation}
Q = 2k \sin(\theta)
\end{equation}
where $2\theta$ is the angle between in the incoming ray and the scattered ray (the scattering angle).

To describe the $Q$-dependence of this scattering, we might consider a Cauchy distribution:
\begin{equation}
\label{eq:LeastSquaresFX}
y \sim \frac{A}{\kappa^2 + Q^2}
\end{equation}
where $\kappa$ is the inverse correlation length: $\kappa=1/r$ and $r$ is the correlation length in the material being studied.  Often, a Cauchy distribution is given as a function of $\gamma$ where $\gamma \equiv \kappa$.  It  is worth noting that the full-width-half-maximum of the Cauchy distribution is equal to $2\kappa$.  $A$ is the amplitude of the scattering, and is related to the contrast terms or spin density if the measurement is calibrated in absolute units.  This is typical of simple critical phenomena in small angle scattering measurements.  The $Q^2$ term is common, as are indeed higher powers of $Q$ for fractal surfaces, even $Q^4$ scattering from smooth surfaces and $Q^6$ from smooth surfaces with significant convolution (a large difference between the mean and gaussian curvatures).  All of these are examples of power law behaviour, which is one of several \emph{long tailed} distributions.  These are widespread in measurements using techniques such as neutron scattering, x-ray scattering, and muon spin relaxation.

In least squares fitting, we begin by assuming the values of $A$ and $\kappa$ are held at fixed values, and calculate the resulting probability of obtaining a neutron in each bin at $Q_i$ for a given $\kappa$, that is $A \times p(Q_i|\kappa)$.  The experimental uncertainties on the $Q$ values are, in this step at least, assumed to be zero.  The resulting probability is compared to a measurement of the probability density, $y_i$, in the form of a count of the number of neutrons in any individual histogram bin $i$.  We then calculate the distance between the model and each data point, $y_i - p(Q_i|\kappa)$.  This may be positive or negative depending on noise and whether the count is above or below the probability function, so we instead work with the square of the distance.  If we were to find a value of $\kappa$ that corresponds to a minimum in the square distance summed over all measurement points, then we find the \emph{least squares estimate} for the parameter $\kappa$.

It is rather trivial to conclude that the size of the histogram bin, $\Delta x$, has some bearing on the determined value of $\kappa$.  If there was only one bin, we would know nothing about $\kappa$ and maybe determine $A$.  As the number of bins increases, we would expect initially to get an improving determination of $\kappa$.  As this process continues, eventually the bin width would become so small that many of the bins have no counts at all.  Since the statistical error on each point is governed by Poisson counting statistics, i.e. for a number of counts in each bin $N_i$, the expected standard deviation of the counts in each bin is $\sqrt{N_i}$, then for small $N_i$ the statistical noise in each bin increases.  The optimum number of bins $n_b$ is then clearly at some yet to be determined  ``medium value'' in the range $2\ll n_b \ll \infty$.

There are several studies that have attempted to optimise the histogram bin width, $\Delta x$.  One of the most well known is the Freedman-Diaconis method \cite{FREEDMAN-DIACONIS}.  The optimum bin width is given by
\begin{equation}
\Delta x = 2 \frac{iqr(x)}{\sqrt[3]{N}}
\end{equation}
where $iqr$ is the interquartile range and $N=\sum_{i}^{n_b}N_i$ is the number of data points counted.  One could of course think of more advanced methods, where the bin width is varied dynamically, and dependent on the local density of counts.  However, this still does not change the fact that there will be windows of integration over areas of $x$ and, as a consequence, some loss of information.

The purpose of this new project, therefore, is to demonstrate a data analysis workflow modelling the probability distributions without computing the histograms at all, i.e. treating the $f(x)$ in the $x$ axis and ignoring $y$ completely in the first instance.  We will also show how to trivially model the amplitudes of the components to $y$, since these are also of scientific interest in the least squares fitting of $y$, to obtain the contrast terms etc.

Before diving into the mathematical treatment, it is worth considering briefly a logical proof of apparent impossibility that will become important later.  One of the most important calibrations of neutron scattering data is the subtraction of an experimental background, which could be random and/or systematic in origin.  Typically, the scientist records a measurement with the sample present, and with the sample absent, and with an efficient absorber at the sample position.  These data sets are used to isolate the contributions to the histogram arising from the sample itself, the sample holder, and the background effects.  It is still important to measure all of these phenomena as part of the event mode data analysis.  We can then argue in a logical sequence that:

\begin{enumerate}
\item \label{objectivenum} We wish to create an analysis workflow that does not rely on histograms or related methods.
\item We require the ability to subtract a measured background from the data.
\item \label{musthistonum} The subtraction of the background requires quantification of the density of events as a function of $x$ of the background and sample contributions, i.e. a histogram or related method.
\item  Points \ref{objectivenum} and \ref{musthistonum} are contradictory.
\end{enumerate}

It seems therefore that our task is logically impossible, but in fact what is needed is a method of quantifying experimental backgrounds that is also not reliant on histograms, along with other corrections.  The method we use in this case is general mixture models.  We will also inject a step in this methodology to use weighted events so that we can correct for solid angle terms, but the same method can be used for detector efficiency and similar systematic effects.

\section{Event Mode Data Analysis}
In this section, we will introduce three methods.  Maximum likelihood estimation (MLE), maximum a posteriori (MAP) using an iterative solver, and finally using Markov Chain Monte Carlo for general numerical sampling of a posterior distribution.  These are all closely related techniques which offer an alternative to least squares fitting.  The basis of these methods is the likelihood function.  Instead of thinking about the relative intensity of the measured particles, $y$, as a function of $Q$ for a given parameter $\kappa$ (as we did in equation \ref{eq:LeastSquaresFX}), the likelihood function answers the question ``If I keep the data $Q$ constant, what is the likelihood that this data came from a parameter distribution described by a parameter $\kappa$?''.  The likelihood function is given by the same Cauchy distribution.  In its normalised form, that is:
\begin{equation}
\label{eq:Cauchy}
f(Q) = \frac{\kappa}{\pi (\kappa^2 + Q^2)}
\end{equation}
Likelihood differs from probability in that \emph{the sum of the likelihoods does not have to equal unity}.  One might consider an extremely large number of candidate distributions, $f(x)$, and compare the relative likelihoods of each to select the most likely model for data analysis.  The sum of all of these likelihoods could be significantly higher than 1.

\subsection{Maximum Likelihood Estimation\label{sec:MLE}}
For a set of $n$ values of many $Q_i$ measured by the instrument, the most likely value for the parameter $\kappa$ --- ``the maximum likelihood estimate'' (MLE) ---  would be given by combining the likelihood terms for each $Q$ value and finding the value of $\kappa$ that makes this likelihood the largest value.  This is a logical ``and'' operation: the total likelihood is the likelihood associated with neutron \#1 \emph{AND} neutron \#2 \emph{AND} $\ldots$ \emph{AND} neutron \#$n$, which means we must multiply the likelihood terms together, so the likelihood is given by:
\begin{eqnarray}
\mathcal{L} & =  &\frac{\kappa}{\pi (\kappa^2 + Q_1^2)} \times \frac{\kappa}{\pi (\kappa^2 + Q_2^2)}  \times \ldots \times \frac{\kappa}{\pi (\kappa^2 + Q_n^2)} \\
   & = & \prod_{i}^n \frac{\kappa}{\pi (\kappa^2 + Q_i^2)}
\end{eqnarray}

The value of $\mathcal{L}$ could get very large or very small very quickly.  With no loss of information, we can flip to a logarithmic encoding of the problem and consider the log-likelihood\footnote{Throughout this article, logarithm is assumed to be the natural log, but it doesn't have to be.} instead:

\begin{eqnarray}
\log(a \times b) & = & \log(a) + \log(b) \\
\therefore \log(\mathcal{L}) & =  & \sum_{i}^n \log\left( \frac{\kappa}{\pi (\kappa^2 + Q_i^2)} \right)
\end{eqnarray}

If we were lucky we could calculate the partial derivative $\frac{\partial \log(\mathcal{L})}{ \partial \kappa}$ and set it to zero, solve for $\kappa$, and validate the maximality by obtaining $\frac{\partial^2 \log(\mathcal{L}) }{ \partial^2 \kappa} < 0$.  Even though it looks like it would have a nice root, a log-likelihood based on a Cauchy distribution actually dosen't.  A gaussian distribution does, and it turns out that you can prove that the maximum likelihood estimate for the centre of the gaussian is the mean of the data points ($\mu = \frac{1}{n}\sum_i^nx_i$), and the width parameter $\sigma$ of the gaussian is the standard deviation of the data points $\sigma = \sqrt{\frac{1}{n}\sum_i^n(x_i - \mu)^2}$.  This is so intuitive to most people they use it without thinking, so well done to everyone who was doing MLE all this time without realising it (like I was!).

Instead, most log-likelihood functions require a numerical approach like Newton iteration.  For a guess of the best parameter value $\kappa$, a better guess is:
\begin{equation}
\label{eq:newton}
\kappa_{+} = \kappa - \frac{\log(\mathcal{L})} { \partial \log(\mathcal{L}) / \partial \kappa}
\end{equation}
Iterate by inserting each $\kappa_{+}$ as a new guess value for $\kappa$ over and over several times until the answer doesn't change much, and you have a candidate solution.  Of course, part of the problem there is figuring out where a good starting point $\kappa$ is so that you actually get the maximum $\log(\mathcal{L})$ and not have the algorithm fly off to infinity trying to find a minimum.  It's not a completely bullet-proof method and you can get it into trouble if you always assume you are getting the right answer.  Often you also need to plot $\log(\mathcal{L(\kappa)})$ to make sure that you are doing something sensible.

\subsubsection{Testing MLE \emph{vs} LSE with Simple Data}

\begin{figure}
\begin{center}
\includegraphics[width=\linewidth]{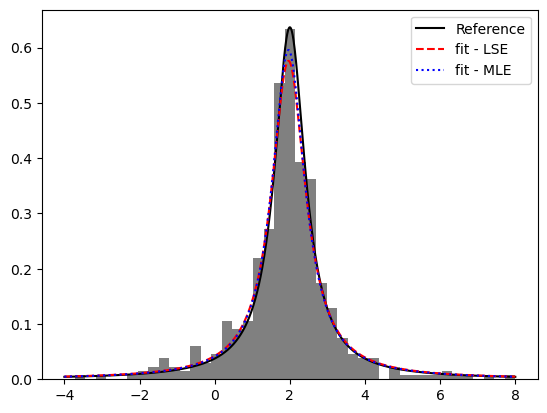}
\end{center}
\caption{Fits to random events, generated by a simple gaussian function, using histograms with least squares regression, and maximum likelihood estimation.}
\label{fig:mle-lse-fits}
\end{figure}

In figure \ref{fig:mle-lse-fits} we show a least squares regression fit of a Freedman-Diaconis-binned histogram and contrast it with maximum likelihood analysis of the same events.  In this case, there is nothing really to distinguish the two fits and either would probably be acceptable for publication.  The MLE line is slightly closer to the reference line, but only slightly.

\begin{figure}
\begin{center}
\includegraphics[width=\linewidth]{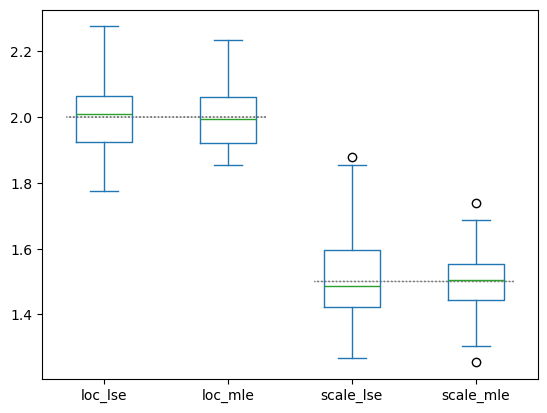}
\end{center}
\caption{Boxplot of the extracted parameters from MLE and LSE from fitting test data similar to that in figure \ref{fig:mle-lse-fits}.}
\label{fig:mle-lse-boxplot}
\end{figure}

In figure \ref{fig:mle-lse-boxplot} we show the extracted parameters from several repeated analyses of random data sets of the same size (500 events).  The true parameter value is indicated by the dotted gray horizontal line.  The mean parameter values for each method are shown, along with their statistical variances (\emph{not} the calculated error bar estimates, but the actual random spread of the obtained values).  We see that for this kind of data, MLE is very slightly superior in accuracy compared to LSE.

\begin{figure}
\begin{center}
\includegraphics[width=\linewidth]{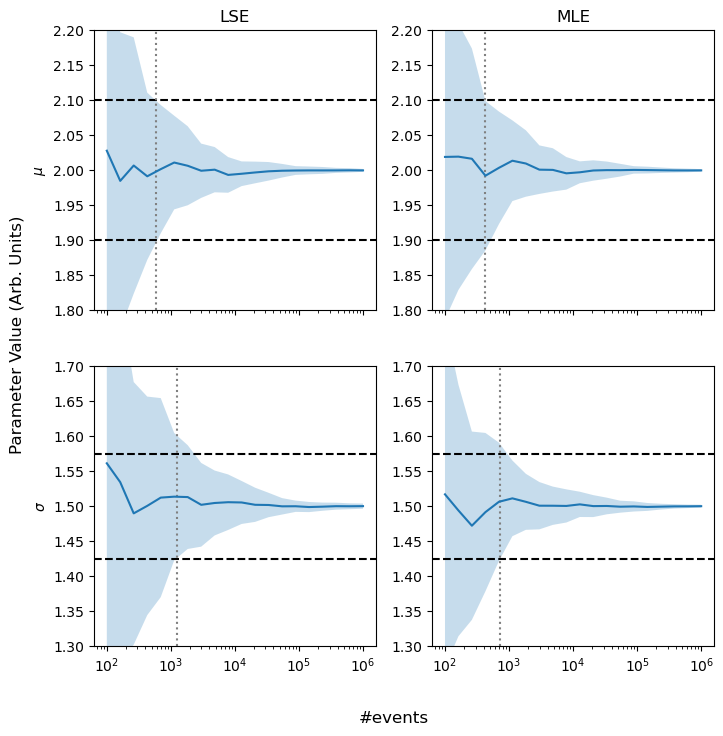}
\end{center}
\caption{Standard deviations and means of the extracted parameters as a function of number of
data events, for both LSE and MLE. Each data point represents the result of 50 analyses of random
samples from the distribution. The dashed horizontal lines represent the boundaries of a region
that is within 5\% of the convergence value. The dotted vertical line is the earliest time that the
standard deviation region is within the 5\% boundary. This indicates that, for both parameters
here, the MLE solution has greater accuracy for a given number of measured data points — it is a
more eﬃcient method.}
\label{fig:mle-lse-convergence}
\end{figure}

One might wonder how the extracted parameter variances evolve as a function of the number of events.  This is revealed in figure \ref{fig:mle-lse-convergence}.  There is indeed a slight, general advantage in using MLE compared to the more traditional LSE method.

\subsubsection{Testing MLE with More Realistic Data}
\begin{figure}
\begin{center}
\includegraphics[width=\linewidth]{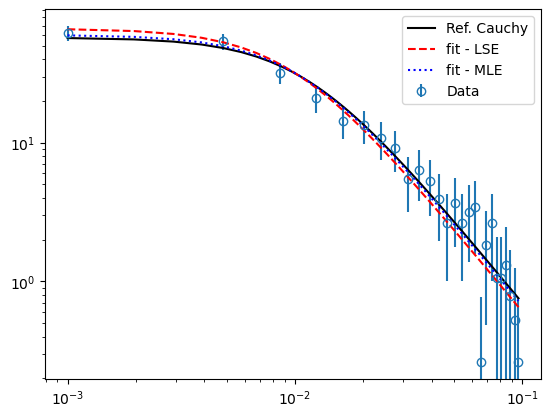}
\end{center}
\caption{Fits to random events, generated by a Cauchy distribution, using histograms with least squares regression, and maximum likelihood estimation.}
\label{fig:mle-lse-sanslike-fits}
\end{figure}

\begin{figure}
\begin{center}
\includegraphics[width=\linewidth]{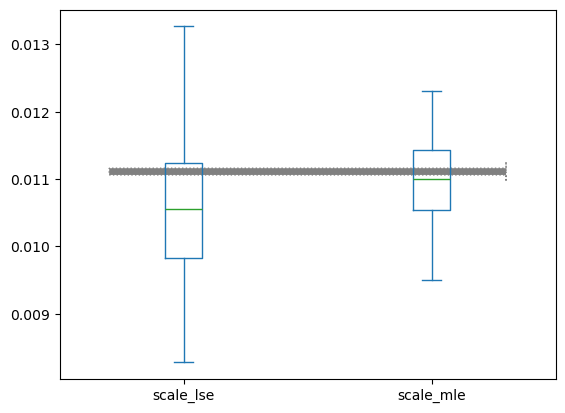}
\end{center}
\caption{Boxplot of the extracted parameters from MLE and LSE from fitting test data similar to that in figure \ref{fig:mle-lse-sanslike-fits}.}
\label{fig:mle-lse-sanslike-boxplot}
\end{figure}

A more SANS-like data set is now tested with the same kind of analysis.  This is shown in figure \ref{fig:mle-lse-sanslike-fits}, again with a least squares regression fit of a Freedman-Diaconis-binned histogram and contrast it with maximum likelihood analysis of the same events.  The differences are once again quite small, but here we can see that the MLE parameterisation may have a tiny advantage in accuracy over LSE.
This is shown better in the boxplot in figure \ref{fig:mle-lse-sanslike-boxplot}.

\subsection{Maximum A Posteriori Estimation \label{sec:MAP}}
It might be the case that we have some prior information about $\kappa$ that helps us narrow our search.  Instead of simply starting with only a guess of $\kappa$ we can also inject our existing knowledge with a prior probability distribution $g(\kappa)$.  $g(\kappa)$ could be a uniform distribution, a fairly broad gaussian, or some other shape depending on what we already know about $\kappa$.  We can then apply Bayes' theorem, which states that:
\begin{equation}
\label{eq:Bayes}
p(A|B) = \frac{p(B|A) p(A)}{p(B)}
\end{equation}
The vertical line means a \emph{conditional probability}, something occurring given that something else has occurred.  $A$ and $B$ are probabilistic events.  Therefore, $p(A|B)$ means ``the probability of event $A$ happening, given that event $B$ has occurred.''  $A$ could represent a value of the parameter we are interested in, such as $k=1$,  and $B$ could be a measurement of a data point, $Q=1$ for example, so the whole equation answers the question ``What is the probability that my parameter equals some value, given that I measured this data point?''

$p(A)$ is the \emph{prior probability}, which describes our knowledge about the parameter \emph{before any measurement takes place}.
$p(B)$ is the \emph{marginal probability}, which describes the overall probability of measuring a data point all things considered, i.e. irrespective of the value of any parameters.  
$p(B|A)$ is the conditional probability of obtaining such a data point given that the parameter has a certain value.  $p(B|A)$ is equal to the likelihood with the two events swapped round, i.e. $\mathcal{L}(A|B)$.  This is what we already hinted at in the text when we introduced equation \ref{eq:Cauchy}, in contrast to equation \ref{eq:LeastSquaresFX}.  It's just the function we are trying to fit, in this case the Cauchy distribution.  $p(A|B)$ is called the \emph{posterior probability}, the updated probability distribution of our parameter, taking into account the event $B$ of us making a measurement that has just occurred.

I have written a fun and trivial example of the application of Bayes' theorem in section \ref{sec:MurderMystery}, if you are new to this maths then feel free to have a read of that before we proceed, and you'll have a better idea of what is going on.

Using equation \ref{eq:Bayes} we can then calculate the posterior distribution of $\kappa$ given the measured points $\mathbf{Q}$:
\begin{eqnarray}
\label{eq:MAPderivation}
p(\kappa | \mathbf{Q}) & = & \frac{p(\mathbf{Q}|\kappa) g(\kappa)}{p(\mathbf{Q})} \\ 
	\label{eq:MAPderivation2}
	& \equiv & \frac{\mathcal{L}(\kappa | \mathbf{Q}) g(\kappa)}{p(\mathbf{Q})}
\end{eqnarray}

The most likely parameter value, given all the data, is the value of $\kappa$ which maximises equation \ref{eq:MAPderivation}.  $p(\mathbf{Q})$ is independent of any parameter values, so we can ignore it as a normalisation term in this particular treatment.  Combining all of these terms over $\mathbf{Q}$ again requires products, which quickly get very large or very small, and we already how to do this so by taking the logarithm of both sides we jump from equation \ref{eq:MAPderivation2} to:

\begin{equation}
\label{eq:mapLL}
\log\left[p(\kappa | \mathbf{Q}) \right] =  \sum_{i}^n \log\left( \frac{\kappa}{\pi (\kappa^2 + Q_i^2)} \right) + \log\left[g(\kappa)\right]
\end{equation}

This is just the same equation that must be maximised for MLE but with an extra log-prior term added to the end.  The Bayesian prior is like a single, subjective data point.  You should be able to see immediately that as your data set gets large ($n \rightarrow \infty$), the prior could become essentially irrelevant and the solutions from MAP and MLE converge.  

One useful way to use the prior to constrain $\kappa$ is to create a $g(\kappa)$ that is essentially unity in all the likely areas (e.g. a piecewise uniform distribution with the first term $g(\kappa)=1$, so $\log(g(\kappa)) = 0$) and essentially zero elsewhere else (a piecewise uniform distribution with the second term $g(\kappa) = 0$, so $\log(g(\kappa))=-\infty$).  Now the value of $\kappa$ is guaranteed to emerge in the area of parameter space that your prior knowledge indicated it would do, because everywhere else no amount of measured data will drag it away from negative infinity.  If the $\kappa$ value you obtained is then right on the edge of that bounding box, then you know you need to relax the $g(\kappa)$ somewhat.  That is a fairly extreme prior, and if the crudeness of this computational baseball bat offends your scientific sensibilities\footnote{This is actually known as ``Cromwell's rule''.  You shouldn't really use a prior with $p=1$ or $p=0$ because what you are really saying is that no amount of evidence could possibly change your mind, and as scientists we should be persuadable if we are given solid evidence.} then you could of course play with gaussians or arctan functions that are smoother and less dramatic, depending on what you already know about the system.  For example, if you are dealing with a powder you could put that under a microscope to determine the particle sizes you can see with visible light.  The mean and standard deviation of those particle sizes could then be used to construct an arctan function that would impose a lower limit on the value of $\kappa=1/r$ for your neutron experiment.  In practice, $\mu$m particles are beyond the resolution of SANS, otherwise people would just use optical microscopes instead of neutrons, but I'm sure you understand the concept.

Basically, \emph{maximum a posteriori is a slightly fancy maximum likelihood estimate with constrained parameters}.

\subsubsection{Parameter Uncertainties from MLE and MAP}
Of course we want to know what the uncertainties are on the parameters.  There is a rather convenient metric that allows us to do this.  The partial derivative with respect to a parameter of the log likelihood is called the \emph{score}.  We've already calculated the score in the maximum likelihood section, because we needed the partial derivatives to do Newton iteration and find the maximum likelihood in the first place.  The Fisher information is an integral of the score squared times the likelihood:

\begin{equation}
\label{eq:FisherInformation}
\mathcal{I}(\theta) = \int_{\rm I\!R} \left( \frac{\partial}{\partial \theta} \log[ \mathcal{L}(x, \theta)] \right)^2 \mathcal{L}(x, \theta) dx
\end{equation}

A thing called the Cram\'er-Rao bound then applies, which means that the lower bound on the uncertainty on the parameter is the inverse of the Fisher information, i.e. $1/\mathcal{I(\theta)}$.   Thus, the Fisher information of equation \ref{eq:FisherInformation} \emph{is the information we are obtaining when we do an experiment}.  It is why we get shrinking parameter uncertainties as we collect more experimental data.

\subsection{Sampling the Posterior with Markov-Chain Monte-Carlo}
Instead of trying to maximise $\mathcal{L}$ and calculate the mode and curvature of the likelihood function, we could instead take linear samples along $\kappa$ and plot the likelihood (I mentioned earlier having to do that sometimes anyway) and take a weighted mean and standard deviation of that function, in other words assuming it was gaussian\footnote{The central limit theorem hints to you that this would probably work even if $\kappa$ was not exactly gaussian distributed.  The Bernstein-von Mises theorem is a more rigorous proof of this: the distribution converges to a gaussian centred at the maximum likelihood estimator, with a covariance matrix that is a function of the inverse Fisher information.}.  However, there are problems with doing that blindly:

\begin{enumerate}
\item What range of the parameter(s) are we going to cover, in other words, the size of the parameter space?
\item How sharp is the distribution of each parameter?  What if you spend a huge amount of time calculating a lot of (almost) zeros?
\item The computational effort expands as $s^k$ where $s$ is the number of points in one axis of the grid and $k$ is the number of parameters you have.  In other words, the dimensionality of your parameter space.  This can rather quickly diverge into a difficult problem to solve.  For example, a 3D spatial reconstruction using maximum entropy (a related technique that is out of the scope of this project) could very well require many gigabytes of RAM and hours of computation time to solve.
\end{enumerate}

Fortunately, this is a known problem.  Back in the 1950s, there were a lot of people trying to simulate nuclear physics ``experiments'' using Monte-Carlo.  They realised that their parameter spaces were: 
\begin{itemize}
\item Large: with lots of dimensions
\item Pointy: with a small interesting part and mostly bad bits everywhere else
\item Slow: difficult or time consuming to evaluate each point
\end{itemize}

These three all apply here.  We might be in the situation of having several terms in the fitting function when we include all random and systematic background effects.  The solution, we hope, is quite well defined with small uncertainty.  We probably want to use this method at a modern instrument with $>10^6$ events per second, and processing all those events could require a GPU in the end.  The solution they came up with is called ``Markov-Chain Monte-Carlo'' (MCMC).  ``They'' is N. Metropolis, A. W. Rosenbluth, M. Rosenbluth, A. H. Teller, and E. Teller.

E. Teller is Edward Teller, famous for creating very large balls of fire.  A. H. Teller is his wife, Ariana.  The Rosenbluths are another married couple who worked in plasma physics.  Metropolis \emph{et. al.}'s idea was essentially the first major advance in variance reduction, which was subsequently generalised by W. K. Hastings, thus it is called the Metropolis-Hastings algorithm.  It was Metropolis who had the idea, and Ariana who wrote the first code.

The algorithm randomly jumps around the parameter space, probabilistically accepting and rejecting move candidates, so that the random samples it generates match those of the underlying probability distribution, whatever that may be.  However, it may take a few hundred samples to ``burn in'' the MCMC algorithm until it stabilises around the maximum and generates good samples.  

The point is this: if our parameter space were to get large and slow to evaluate, for example in the form of a high-dimensional log-likelihood function with many events, we can simplify our problem and sample it with MCMC.  Then, to determine the parameter values and their uncertainties, we just need to compute the mean and the standard deviation of the MCMC samples along the particular direction of interest, thanks to the central limit theorem.  One note of caution: if the parameter space is multimodal, MCMC could become trapped by a local optimum.  It is wise to plot out the samples to see that you really are getting an almost gaussian, unimodal sampling output for each parameter.

It's worth noting that real world data is noisy, which makes the parameter distributions not cleanly gaussian, but a sharp-ish gaussian sitting on a long-tailed mess.  The mean and standard deviation trick might not work in that scenario because the background parts are non-uniform and non-symmetrical with respect to the peak centre.  What works there is finding the mode of the distribution instead (so yes, we have to histogram the parameter) and we can get the most prominent peak position and its standard deviation using signal processing from the scipy library.

In any case, we are now touching on \emph{probabilistic programming} where we assume that each parameter is drawn from a distribution, and instead of trying to find a parameter value we are trying to establish a mode and standard deviation for each parameter.  There are a bunch of python packages that do this probabilistic programming $+$ MCMC combo, and we tested several of the major ones.  These are (along with a little commentary):

\paragraph{PyMC (Rejected)} 
It has a nice, clean API that is easy to understand, and the option to use many MCMC samplers.  But it was very flaky to set up and run due to dependency problems.  As far as I can tell, it's based on Theano which has been discontinued, and this is what seems to be causing said dependency problems.  Of all of the tested packages, this was my favourite API.  I think it has all the distributions from Scipy, which is plenty.  You can make it run for isolated example cases but it wasn't deployable for general use.  I think in some scenarios I was even getting import errors for numpy depending on what backend I was trying to fire up.  The workaround was to have a separate conda environment for every type of problem.  Not good.

\paragraph{Pomegranate (Rejected)}
This is probabilistic programming, based on pytorch.  It is the lightest API of all of them, but the library of probability distributions is very small, making it unusable for now.

\paragraph{Tensorflow Probability (An option)}
This one is self explanatory.  It relies on JAX (which I had trouble with on apple silicon and PyMC), but other back ends are supported.  It might be a solid platform to deploy on in the long term.

\paragraph{Edward (An option)}
It looks promising, but the API seemed a lot more involved than some of the others.  It also looks like the all the scipy distributions are available which is a good sign.

\paragraph{EMCEE (Tested and deployable)}
This is just a good, barebones MCMC sampler, so you have to code up everything else yourself.  This is actually the route I went down, because then you know exactly where you are and there are no black boxes when it comes to the important maths.  It also runs on almost anything with zero drama.  The next section explains how this was built.  Actually, EMCEE \cite{EMCEE} is better than good.  There are significant advantages over the simpler Metropolis-Hastings algorithm due to some work by Goodman and Weare \cite{GOODMAN-WEARE}, the EMCEE team basically took the Goodman-Weare work and coded it up in python, plus added a few innovations of their own.

\subsubsection{Implementation of MCMC-Sampled Bayesian Analysis}

Back in equation \ref{eq:mapLL} from the maximum a posteriori method in section \ref{sec:MAP} we had Bayesian updating from a prior for a single probability distribution.  This is a ``zero background'' test because it only has one source of neutrons: the sample.  It turns out that even so, it is surprisingly resistant to background effects.  I tested it with a signal to noise ratio of 1:1 and it still produced results with an accuracy of within 10\%.  Most of the instruments are seeking something like $10^6$:1.  Nonetheless, we are not done.  To do event mode data analysis we still have to demonstrate:

\begin{enumerate}
\item \label{point:GMM} Multiple contributions, e.g. background, sample holder...
\item \label{point:Weights} Event weights, for things like solid angle corrections, detector efficiency...
\end{enumerate}

Problem \ref{point:GMM} is solved by building a \emph{general mixture model}.  Let us assume that each neutron event, $Q_i$, could be generated by either of two probability distributions.  First, we have the same Cauchy distribution as demonstrated in equation \ref{eq:mapLL} that comes from the interesting science.  The second distribution we will assume is a flat background effect, i.e. a uniform distribution u(Q) where each $Q$ is equally probable.  With least squares fitting we could subtract the background, but we cannot do that here because to do so would require some kind of histogram, so instead we must put it in the model.  Each neutron event has a parameter $Z_i$, where $Z \in [0,1]$.  Think of $Z_i$ as being like a switch parameter that says whether the event comes from the sample ($Z_i=1$) or the background, ($Z_i=0$).  That's a lot of parameters, one for each data point in fact.  Since this is a logical \emph{OR} operation, in the language of mathematical probability functions that will be encoded as an addition operator; whilst a logical \emph{AND}  operation is encoded as a multiplication operator.  Either the neutron comes from the sample \emph{and} it is defined by a Cauchy distribution \emph{or} the neutron comes from the background \emph{and} it follows a uniform distribution.  The total likelihood is a logical combination of the information from neutron \#1 \emph{and} neutron \#2 \emph{and} $\ldots$ \emph{and} neutron \#{}$n$.  Our likelihood function has therefore now evolved into this:

\begin{equation}
\label{eq:gmmRaw}
\mathcal{L} =  \prod_{i}^n \left[ Z_i \times \frac{\kappa}{\pi (\kappa^2 + Q_i^2)} + (1-Z_i) \times u (Q_i) \right]
\end{equation}

In more rigorous language, what we have just done is multiply by a prior $p(Z_i)$ for the categorisation of the data point $Q_i$ where the prior is defined by:
\begin{equation}
\label{eq:MixPrior}
p(Z_i) =  \begin{cases}
	M  & \text{if}\,Z_i = 0 \\
	1-M & \text{if}\,Z_i = 1
	\end{cases}
\end{equation}
Next, we can do something that looks like some kind of black magic, but is based on a paper by Hogg \cite{HOGG-BAYESIAN} and a website by Foreman-Mackey \cite{DFM-WEBSITE}.  We are going to integrate over the $Z_i$ values to make the parameter space smaller.  This is called ``marginalising out'' the parameters $Z_i$ and, since the $Z_i$ are all discrete variables the integral therefore becomes a sum (and the sum and product orders can be swapped around \emph{iff} the $Q_i$ are all independent --- which they are):

\begin{eqnarray}
\label{eq:gmm1}
\mathcal{L} & = &  \sum_{Z_i} \prod_{i}^n \left[ Z_i \frac{\kappa}{\pi (\kappa^2 + Q_i^2)} + (1-Z_i) u (Q_i) \right] \\
 		& = & \prod_{i}^n M \frac{\kappa}{\pi (\kappa^2 + Q_i^2)} + (1-M) u (Q_i)
		\label{eq:gmm2}
\end{eqnarray}

$M$ is the mean mixing fraction of the sample signal.  $M=1$ means that the relative background level is negligible.  $M=0$ means that there is no detectable signal and only background.  We can then use MCMC to sample a parameter space with only one extra dimension, namely a space of $(M, \kappa)$.  Equation \ref{eq:gmm2} feels intuitive, and you might wonder why we didn't just start with that and have done with it.  The reason for going through the marginalisation steps above is to link all of this back to Bayes' theorem on an event-by-event basis, and show that it doesn't just appear out of thin air.

When we code this up, we still have to constrain $M$ to fall in the range $0<M<1$ so there is still an explicit prior in the code even though it doesn't appear in the maths of equation \ref{eq:gmm2}, but in equation \ref{eq:MixPrior} instead.  Our MCMC algorithm will then converge on a point $(M, \kappa)$ that maximises equation \ref{eq:gmm1}-\ref{eq:gmm2} and we get the relative sample ($M$) and noise ($1-M$) mixing weights  plus $\kappa$, along with their standard deviations, just as we would try to get from a least squares fit with sample and noise terms.

It is trivial to generalise equation \ref{eq:gmm2} to more terms, by turning $p(Z_i)$ into a higher dimensional simplex, so you end up with a set of mixing parameters that is one fewer in number than the number of terms in the probability distribution that you are fitting, and as usual you use the prior to enforce that each mixing parameter spans the range 0--1 and that the sum of the mixing terms is unity.

If we are interested in contrast, calibration to absolute units etc, we have the mixing weights for the distributions and the total number of neutron events.  It's therefore trivial to compute the amplitude parameter $A$ from equation \ref{eq:LeastSquaresFX} and it's uncertainty, if this is important to somebody.

Our final step is to deal with problem \ref{point:Weights}, namely the weighting of individual data points $Q_i$ to take into account detector efficiency, solid angle corrections, etc.

If we flip equation \ref{eq:gmm2} into a logarithmic space, we get:
\begin{equation}
\label{eq:loggmm2}
\log(\mathcal{L}) = \sum_{i}^n \log \left[ M \frac{\kappa}{\pi (\kappa^2 + Q_i^2)} + (1-M) u (Q_i) \right]
\end{equation}
and if we were to apply a statistical weight $w_i$ to each event $i$ this becomes:
\begin{equation}
\label{eq:logwgmm2}
\log(\mathcal{L}) = \sum_{i}^n w_i \times \log \left[ M \frac{\kappa}{\pi (\kappa^2 + Q_i^2)} + (1-M) u (Q_i) \right]
\end{equation}

Remembering another logarithmic identity
\begin{equation}
a \log(b) = \log(b^a)
\end{equation}
we might worry that in linear space we are raising the likelihood of each event $i$ to a power $w_i$.  We are!
\begin{equation}
\label{eq:linweight}
\mathcal{L} = \prod_{i}^n \left( M \frac{\kappa}{\pi (\kappa^2 + Q_i^2)} + (1-M) u (Q_i) \right)^{w_{i}}
\end{equation}
If $w_i=0$ then the event should contribute no information to the analysis process.  $a^0=1$ so, in the product of equation \ref{eq:linweight}, the event multiplies the log-likelihood of the other events by unity and so changes nothing.  In equation \ref{eq:logwgmm2} of course we are multiplying the event by zero so the event contributes nothing to the sum.  If $w_i=1$ then the event should contribute a full unit of information to the analysis process.  $a^1=a$ so, in the product the event does indeed multiply the log-likelihood of the other events by its full log value, and likewise in the sum it adds its full value to the linear likelihood.

\subsubsection{Testing of MCMC-Sampled Bayesian Analysis}
\begin{figure}
\begin{center}
\includegraphics[width=\linewidth]{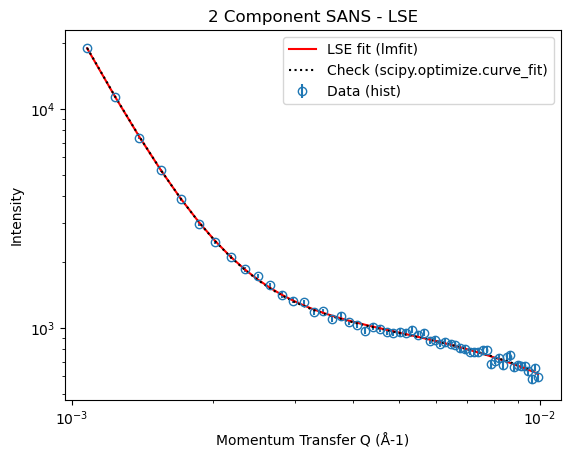}
\end{center}
\caption{Simulation of SANS-like reduced data set, histogrammed according to the optimal Freedman-Diaconis method.  The solid line is a least-squares fit to the histogram, using the python pacakge ``lmfit'', alongside a manual invocation of scipy.optimize.curve\_fit to validate the methodology.}
\label{fig:twocomponent-lse-fit}
\end{figure}

In this demonstration, we will generate a synthetic SANS spectrum with both a signal and a large background.  Figure \ref{fig:twocomponent-lse-fit} shows a traditional analysis of this data.  There is a critical scattering component following the common Cauchy distribution, and a Porod-like $\propto Q^{-4}$ systematic background.  The relative intensity of the two components is 1:1.  The fit to this data is reasonable and probably would not have any challenges in terms of quality of work.

\begin{figure}
\begin{center}
\includegraphics[width=\linewidth]{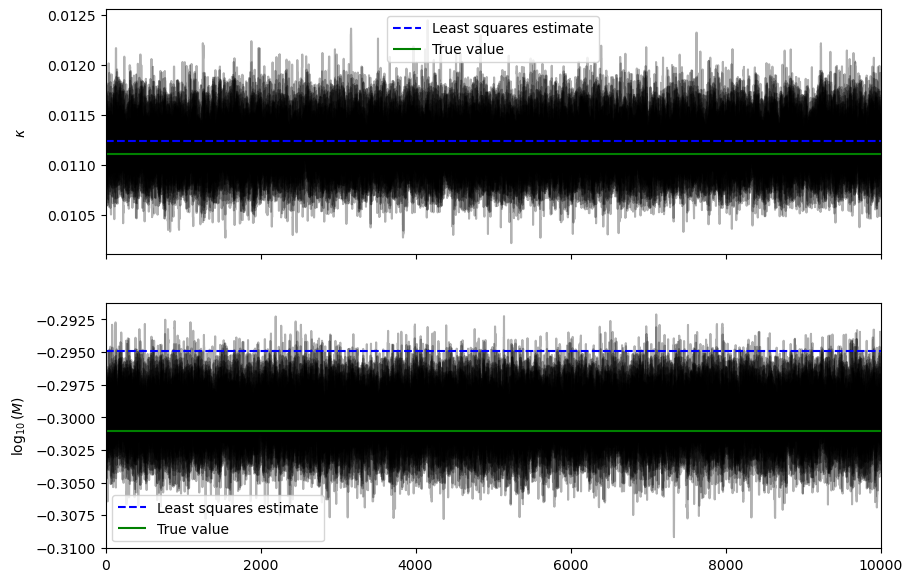}
\end{center}
\caption{10,000 samples of MCMC from the $(M,\kappa)$ parameter space, using 32 parallel random walkers.}
\label{fig:MCMC-samples}
\end{figure}

If we now switch to event mode, and use the MCMC-sampled Bayesian method, the samples of the $(M, \kappa)$ parameter space are shown in figure \ref{fig:MCMC-samples}.  The true parameter value is shown as a solid horizontal line, and we can see that the MCMC samples are pretty close.  On the other hand, the least squares estimate is not as good, at least for the mixing parameter.  It is known that long-tailed distributions sometimes exhibit systematic effects with least squares fitting \cite{CLAUSET-POWER-LAW}, and this is probably a mild example of this.

\begin{figure}
\begin{center}
\includegraphics[width=\linewidth]{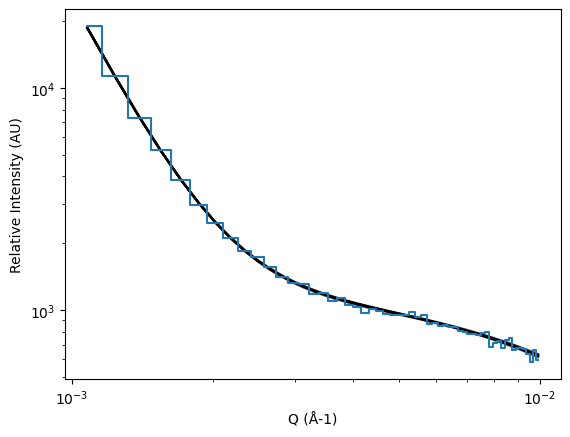}
\end{center}
\caption{Comparison of the PDF resulting from the mode of each of the MCMC walker parameter distributions (solid smooth lines) with a histogram of the data points (solid step line).}
\label{fig:MCMC-fit-histo}
\end{figure}

There are two methods one might choose to plot a ``fit'' of the data from the MCMC analysis, in other words, compute the PDF of the intensity vs $Q$ based on the extracted parameters from the MCMC analysis and scale it so that the integral is the same as the histogrammed data points.  The first way is to plot the PDF coming from the parameter value associated with each of the parallel walkers.  The advantage of this method is that any variance of the parameters is visible alongside the measurements.  This is shown in figure \ref{fig:MCMC-fit-histo}, where the lines overlap along the left side of the graph and spread out slightly to the right side.
The second method is to take the mean or mode of each parameter distribution, and plot a single PDF curve with those values.

As for the histogram method itself, an alternative is \emph{kernel density estimation} (KDE) \cite{Hardle2004-hh}.  In this method, we convolve a kernel function, $g(Q)$, with the density of measurements $Q_i$ and plot the resulting numerical integral on a reference set of axis points $Q_j$.  The $Q_j$ might be, but do not have to be, the same as the central points for each histogram bin, but doing so makes the comparisons simpler to code.  The function $g(Q)$ could be a Gaussian or some other shape, but the most important point is that the graph is created via a sliding integral rather than a fixed grid of integral boxes.  There are a lot of different KDE libraries and kernel functions one can use.  Here, we tested the scikit-learn \cite{SCIKIT-LEARN} implementation of KDE with an Epanechnikov kernel function, and the Gaussian KDE method in scipy \cite{SCIPY}, using a manually-determined bandwidth on a case by case basis.  The motivation for doing this is twofold:
\begin{enumerate}
\item To eliminate any coding errors dealing with histogram binning widths, centres, along with psychological or aesthetic distractions
\item To eliminate any potential systematic effects from the analysis chain, such as those mentioned earlier after Clauset \cite{CLAUSET-POWER-LAW}.
\end{enumerate}

\begin{figure}
\begin{center}
\includegraphics[width=\linewidth]{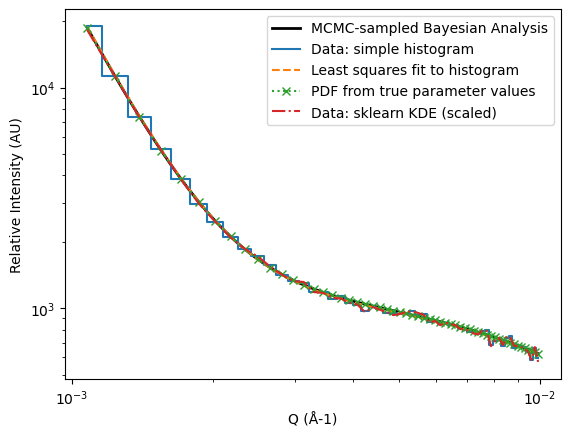}
\end{center}
\caption{A least-squares fit of histogrammed data, compared to a MCMC-sampled Bayesian analysis and KDE plot, along with the PDF associated with the parameters actually used to generate the events.}
\label{fig:MCMC-fit-comparison}
\end{figure}

Figure \ref{fig:MCMC-fit-comparison} overlaps histogram of the data set, the least squares fit, a single PDF from the mode of the MCMC parameters, the expected PDF from the true parameter values, and a KDE curve of the data set: so that we can establish a solid starting point, where all the methods are essentially in agreement with a known parameter distribution and already in a somewhat challenging case with a long tailed distribution.

\section{Real World Data Analysis}
The academic paper version of this report includes analysis of real-world ARCS data.  This section is skipped here because the method is already described above, and the purpose of this report has changed.  Readers are invited to find that paper/preprint submitted in 2026.

\section{Conclusions}
We have demonstrated a weighted event mode data analysis workflow.  The proposed method, a Bayesian Markov-Chain Monte-Carlo analysis, is orders of magnitude more efficient than least squares, and avoids some potentially problematic biassing for some types of probability distribution that are common in neutron scattering \cite{CLAUSET-POWER-LAW}.

\newpage
\section{Appendix: A Light-Hearted Murder Mystery\label{sec:MurderMystery}}

In an actual murder investigation, we would ignore the absolute probabilities and concentrate on the relative likelihoods, but in this example it's really instructive to consider the marginals, particularly in light of real investigations that have gone badly wrong and put the wrong people in jail for crimes they did not commit.

Dr Black has been murdered at his big house during a dinner party.  The inspector arrives and does a DNA analysis of the murder weapon (the candlestick) at the scene of the murder (the conservatory).  He's feeling pretty good, because he only has to figure out the person and he's not even left the first room yet.

The initial suspicion is equally distributed, as shown in table \ref{tab:initialSuspicion}.

\begin{table}[h!]
  \begin{center}
    \caption{Initial suspicion at Dr. Black's house.}
    \label{tab:initialSuspicion}
    \begin{tabular}{l|l} 
      \textbf{Suspect} & \textbf{Probability} \\
            \hline
	Miss Scarlett & 0.167 \\
	Col. Mustard & 0.167 \\
      Mrs White & 0.167 \\
      Rev. Green & 0.167 \\
      Miss Peacock & 0.167 \\
      Prof. Plum & 0.167
    \end{tabular}
  \end{center}
\end{table}

The DNA results come back that evening, and Miss Scarlett is a match!

Recall equation \ref{eq:Bayes}:
\[
p(A|B) = \frac{p(B|A) p(A)}{p(B)}
\]
For the DNA test, we have an evidence match $B\equiv m$, and guilt for the act of murder $A\equiv G$:
\[
p(G|m) = \frac{p(m|G) p(G)}{p(m)}
\]

The prior $p(G)$ is equal suspicion for all suspects, as given in table \ref{tab:initialSuspicion}.  Strictly speaking there should also be an entry for ``someone else'' but we'll ignore that for this fun exercise.

$p(m|G)$ is the likelihood that the DNA would match assuming Miss Scarlett is indeed guilty of the murder.  The DNA kit claims that in the general population, the test would return a match between two randomly chosen people with a probability $p(FP)=0.07^5$, which is roughly 2:1\,000\,000.  This is basically the false positive rate ($FP$) because it's the probability that you get a DNA match purely by chance.  \emph{A lot} of people get impressed by these small numbers, and are lead to believe if it's wrong with odds 2:1000\,000 then a positive match is correct with odds 1000\,000:2.  The inspector here thinks that the probability of her guilt is $>$99.99\% given that she has DNA on the murder weapon.

I warn you there is a huge mistake here, but a first glance with Bayes' theorem in a bit more detail might lead you to believe the inspector is correct on his hunch.  The marginal probability $p(m)$ is the probability that we would get such evidence whether or not the suspect is guilty, so there are two terms: the probability that her DNA would be on the weapon if she were guilty, plus the probability that her DNA would be on the weapon if she were innocent.  The first term is actually the same as the numerator, i.e. the probability she has a positive test result assuming she is the murderer, $p(m|G)$, multiplied by (\emph{$\equiv$AND}) the prior probability she is the murderer $p(G)$.  The second term is the probability that the test is a match assuming she is innocent, $p(m|\bar{G})= p(FP)$ multiplied by the probability she is innocent $p(\bar{G}) = 1-p(G)$.

\begin{eqnarray}
\label{eq:BayesWrong}
p(G | m) & = & \frac{p(m|G) \times p(G)}{p(m)}  \\
 & = & \frac{p(m|G) \times p(G)}{p(m|G) \times p(G) + p(m|\bar{G}) \times p(\bar{G})}  \\
 & = & \frac{0.999998 \times 0.167 }{0.999998 \times 0.167 + 0.0000017 \times (1-0.167)}  \\
 & \approx & 99.999\%
\end{eqnarray}

Things don't look good for Miss Scarlett.  Fortunately, Prof. Plum is on hand to explain one or two facts of the matter, and inject a little reality, and remove a bit of marketing hype from the lab equipment vendor \cite{BENTLEY-BAYES-COVID}.

Firstly, he points out that the true positive $p(TP) \neq 1-p(FP)$.  Given that a planned murder would probably involve the guilty party trying to clean the crime scene and wear gloves, he suggests that the probability of the murderer leaving a trace should be much lower, more like 0.8.

Secondly, he points out that the false positive rate \emph{in the field} is much higher than $p(FP)=0.07^5$.  It depends on the lack of contamination in a whole chain of events, proper labelling of sample bags, proper cleaning of lab equipment etc.  He suggests $p(FP)=0.05$ and even that is being generous, he thinks.  On the blackboard, he then computes:
\begin{eqnarray}
\label{eq:BayesRight}
p(G | m) & = & \frac{p(m|G) \times p(G)}{p(m)}  \\
 & = & \frac{p(m|G) \times p(G)}{p(m|G) \times p(G) + p(m|\bar{G}) \times p(\bar{G})}  \\
 & = & \frac{0.8 \times 0.167 }{0.8 \times 0.167 + 0.05 \times (1-0.167)}  \\
 & \approx & 76\%
\end{eqnarray}

The inspector is still optimistic, but then Prof. Plum takes samples all around the house and they all come back positive for Miss Scarlett.  It turns out she was romantically involved with Dr. Black and spent a lot of time in the house.

The inspector then states that only 13\% of murders are committed by strangers, and becomes very interested that Rev. Green is the victim's cousin.  He calculates a guilt probability for Rev. Green:
\begin{eqnarray}
\label{eq:BayesBrother}
p(G | m) & = & \frac{p(m|G) \times p(G)}{p(m|G) \times p(G) + p(m|\bar{G}) \times p(\bar{G})}  \\
 & = & \frac{0.87 \times 0.167 }{0.87 \times 0.167 + 0.13 \times (1-0.167)}  \\
 & \approx & 57\%
\end{eqnarray}

Prof. Plum then points out that the above equation applies to pretty much everyone there, since they all knew Dr. Black, except Col. Mustard who was Prof. Plum's tag-along guest.  So, he suggests that instead of the above, we could calculate the inverse and downgrade the guilt for Col. Mustard:
\begin{eqnarray}
\label{eq:BayesMustard}
p(G | m) & = & \frac{p(m|G) \times p(G)}{p(m|G) \times p(G) + p(m|\bar{G}) \times p(\bar{G})}  \\
 & = & \frac{0.13 \times 0.167 }{0.13 \times 0.167 + 0.87 \times (1-0.167)}  \\
 & \approx & 3\%
\end{eqnarray}

When asking who was where when the murder took place, we find that Prof. Mustard remembers talking to Rev. Green in the dining room; and Miss Scarlett was talking to both Mrs Peacock and Prof. Plum in the drawing room.  He wants to be generous with our rate of mistakes in who-where-when problems and remembering faces (police line ups etc) because these are not 100\% accurate like in the movies.  In fact, they are notoriously wrong.  Sometimes entire statements and quotes are attributed to entirely the wrong people in our memories.  Even whole evening events and vacations can place the wrong people.  He suggests some loose boundaries of $p($correct$)=0.9$ and $p($wrong$)=0.1$ for recollecting people being in a location, i.e. providing an alibi statement \footnote{Dr. Black has no CCTV and doesn't issue RFID tags to his guests.}.  We can update each person's probability in the same way as demonstrated in the equations above.  The results are shown in table \ref{tab:finalSuspicion}.
\begin{table}[h!]
  \begin{center}
    \caption{Evolution of suspicion at Dr. Black's house as evidence is assessed.}
    \label{tab:finalSuspicion}
    \begin{tabular}{l|r|r|r|r} 
      \textbf{Suspect} 	& \textbf{Prior} 	& + \textbf{Known} 	& + \textbf{Dining Rm} 	& + \textbf{Drawing Rm}\\
            \hline
	Miss Scarlett 	& 0.167 		& 0.194 			& 0.24				& 0.034 \\
	Col. Mustard 	& 0.167 		& 0.029			& 0.003				& 0.003\\
 	Mrs White 		& 0.167 		& 0.194			& 0.24 				& \textbf{0.87} \\
      	Rev. Green 		& 0.167 		& 0.194			& 0.026 				& 0.03\\
     	Miss Peacock 	& 0.167 		& 0.194			& 0.24 				& 0.03\\
     	Prof. Plum 		& 0.167 		& 0.194			& 0.24 				& 0.03\\
    \end{tabular}
  \end{center}
\end{table}

At this point, your brain is already telling you intuitively that it was probably Mrs White who should be investigated, since she has no alibi.  We see that our intuitive confidence level is supported by the maths.  The probability of her guilt is almost 90\%, even with these rather vague experimental accuracies for evidence gathering.  This is the beauty of Bayes' theorem, the incremental addition of several layers of evidence still drives the result towards the right conclusion, even when dealing with tests that are not that accurate.  In fact, because the accuracy of the tests is taken into account, the absolute test accuracy is not that relevant.  It's a bit like using ZFS on cheap hard drives, the maths takes care of things.  As a side-note, the converse is also true, which was the whole point of my paper \cite{BENTLEY-BAYES-COVID}, that if you only use a single cheap test with wooly error rates \emph{in the field}, then assume the beautiful \emph{laboratory} error rates, and then integrate over a population of millions of people, then a lot of what you are seeing might just be random noise.  You might also give people a false sense of security by ``clearing'' people who actually present a real danger to society (false negative rate is higher in the field than assumed laboratory rates), or create a scenario where an innocent person has to prove that they are not-guilty of an accusation much like in F. Kafka's novel \cite{KAFKA-TRIAL} (false positive rate is higher in the field than assumed laboratory rates).

\section{Appendix: Finding Lost Ships, Planes, and Things}
This section shows how to find things using Bayes' theorem.  Lets say a boat sank in the ocean but you don't know exactly where.  You can create a grid of search boxes of equal size, say $1\times 1$\,km$^2$.  The prior probability of finding the boat could be calculated using a whole slew of expert input.  Then you can layer on radio history, radar signals, phone records etc, and then you go to the square with the highest probability of having the boat in it.  Lets say the probability that the boat is in the box is $p$, and assuming the boat is there, the probability of successfully finding it is $q$.  $q<1$ because if the water is very deep, for example, maybe you miss it.  Even under perfect conditions, maybe just at that moment when you pass the boat you are distracted by a seagull, or a phone call, or a radio conversation.  In practice, $q\sim f(d)$ where $d$ is the water depth, but lets just have a fixed $q$ for now.

Lets first consider how to update if we search the box and do not find the boat.  We use Bayes' theorem of course, and consider the event that the boat is there, $B$, in light of acquired data, $d$, that the boat was not found there.  The converse situation is that the boat is not in this box, and of course the probability then of not finding it $p(d|\bar{B})=1$.  The marginal probability of obtaining the data $p(d)$ is formulated as either the boat is there and we didn't find it, or the boat is not there (and we still didn't find it, of course).

\begin{eqnarray}
p( B | d) & = & \frac{p(d|B) p(B)} {p(d)}  \\
              & = & \frac{p(d|B) p(B)} {p(d|B) p(B) + p(d|\bar{B}) p(\bar{B})}  \\
\therefore p_{new} & =  & \frac{ \bar{q} p }{ \bar{q} p + \bar{p} } \\
              & = &  \frac{ (1-q) p }{ (1-q) p + (1-p)} \\
              & = & p \frac{ (1-q) }{ 1-pq}
\end{eqnarray}

Considering the perspective of a neighbouring box, we update the probability $r$ of all the other boxes in the same way.  In this case, the data $d$ is still not finding the boat in the first box, it's the only data point we have.  Event $\bar{B}$ is that the boat is not located in the search box.  The prior is $r$, and the likelihood $p(d|\bar{B})=1$ because the boat is not in the box that was searched above.  The terms in the marginal are the same because they haven't changed.  So we have:
\begin{eqnarray}
p( \bar{B} | d) & = & \frac{p(d|\bar{B}) p(\bar{B})} {p(d)}  \\
\therefore r_{new} & =  & \frac{ 1 \times r }{1-pq } \\
\end{eqnarray}

If you are interested in this kind of approach to practical problems, the same kinds of searches are used to find ships and planes that go missing.  See, for example, this wikipedia page: \\
\href{https://en.wikipedia.org/wiki/Bayesian\_search\_theory}{https://en.wikipedia.org/wiki/Bayesian\_search\_theory}

For fun, I once made a map of Europe and did a Bayesian search for the lost city of Atlantis, which was incrementally updated using statements on its location from the stories, one by one.  I fed these in using the probability that witness statements are wrong around 30\% of the time.  This might sound surprising, but it's actually true.  If you survey an educated society about their beliefs, around 30\% of the people believe things that are demonstrably false.  Anyway, once you crunch all the numbers, the map shows you the same logical conclusion that agrees with current academic consensus, which is that whilst there are some \emph{relatively} interesting locations scattered around southern Europe, and a peak probability in the south of Spain, the absolute probability of these locations being Atlantis is low - about 1\%.  The story has a 99\% probability of being a myth based on a number of historical events that were geographically separate.

\begin{figure}
\begin{center}
\includegraphics[width=0.8\linewidth]{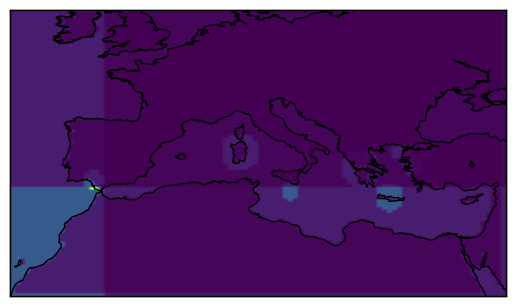}
\caption{Bayesian search for the location of the lost city of Atlantis!  The absolute probability value of the peak in southern spain is 0.01\%, indicating that there is no actual location for the city.  It is probably a myth based on several locations spread around the mediterranean area.}
\end{center}
\label{fig:atlantisMap}
\end{figure}

\bibliography{eventMode}
\bibliographystyle{unsrt}

\end{document}